\documentstyle[11pt]{article}

\begin{document}

\newtheorem{thm}{Theorem}[section]
\newtheorem{lemma}[thm]{Lemma}
\newtheorem{defin}[thm]{Definition}
\newtheorem{rem}[thm]{Remark}
\newtheorem{cor}[thm]{Corollary}

 %
 %
 %
 %

\def\fn{ \baselineskip = 0pt
\vbox{\hbox{\hspace*{3pt}\tiny $\circ$}\hbox{$f$}} \baselineskip = 12pt\!}
\def\gn{ \baselineskip = 0pt
\vbox{\hbox{\hspace*{2pt}\tiny $\circ$}\hbox{$g$}} \baselineskip = 12pt\!}
\def\lap{\bigtriangleup}
\def\be{\begin{equation}}
\def\ee{\end{equation}}
\def\bea{\begin{eqnarray}}
\def\eea{\end{eqnarray}}
\def\beas{\begin{eqnarray*}}
\def\eeas{\end{eqnarray*}}

\def\dt{\partial_t}
\def\dx{\partial_x}
\def\dv{ \partial_v }
\def\R{{\rm I\kern-.1567em R}}
\def\N{{\rm I\kern-.1567em N}}
\def\Z{{\sf Z\kern-.3567em Z}}
\def\s{\sigma}
\def\supp{\mbox{\rm supp}}
\def\div{\mbox{\rm div}}
\def\n#1{\vert #1 \vert}
\def\nn#1{\Vert #1 \Vert}

\def\prf{\noindent
         {\em Proof :\ }}
\def\prfe{\hspace*{\fill} $\Box$

\smallskip \noindent}

\title{Global existence of classical solutions to the Vlasov-Poisson system
       in a three dimensional, cosmological setting}
\author{Gerhard Rein\\
        Mathematisches Institut der Universit\"at M\"unchen\\
        Theresienstr.\ 39, W8000 M\"unchen 2, Germany\\
        and \\
        Alan D.~Rendall\\
        Max-Planck-Institut f\"ur Astrophysik\\
        Karl-Schwarzschild-Str.\ 1, W8046 Garching, Germany}
\date{}
\maketitle
\begin{abstract}
The initial value problem for the Vlasov-Poisson
system is by now well understood in the case of an isolated system
where, by definition, the
distribution function of the particles as well as the gravitational
potential vanish at spatial infinity. Here we start with homogeneous
solutions, which have a spatially constant, non-zero mass density and
which describe the mass distribution in a Newtonian model
of the universe.
These homogeneous states can be constructed explicitly,
and we consider deviations from such
homogeneous states, which then satisfy a modified version of the
Vlasov-Poisson system. We prove global existence and uniqueness of
classical solutions to the corresponding initial value problem
for initial data which represent spatially periodic deviations
from  homogeneous states.
\end{abstract}
 %
 %
 %
 %
\section{Introduction}
Consider a large ensemble of particles which interact only by the
gravitational potential which they generate collectively.
If the particles are described by a time-dependent distribution function
$f=f(t,x,v)$ on phase space and if we neglect relativistic as well as
collisional effects, then the time evolution of the
ensemble is governed by the Vlasov-Poisson system
\be \label{vl}
\dt f + v\cdot \dx f - \dx U \cdot \dv f =0,\
\ee
\be \label{po}
\lap U = 4\pi \rho,
\ee
where
\be \label{rho}
\rho (t,x):= \int_{\R^3} f(t,x,v)\,dv
\ee
denotes the spatial mass density generated by $f$,
$U=U(t,x)$ denotes the gravitational potential, and
$t\geq 0$ denotes time, $x\in \R^3$ position, and $v\in \R^3$ velocity.
The same system---with a minus in front of the right hand side of Poisson's
equation---models a plasma of electrons or ions which interact by
electrostatic Coulomb forces. Here we are concerned only with
the stellar dynamic case.

In the corresponding initial value
problem one prescribes the phase space density at $t=0$, i.\ e.\
$f(0)=\fn$ is a given, nonnegative function on $\R^6$. However, stated in this
form, this initial value problem is ill-posed. It has to be
supplemented with a boundary condition for the potential $U$
at spatial infinity. If one wants to model an isolated system,
for example a galaxy in an otherwise empty universe, one requires that
\be \label{bc}
\lim_{x \to \infty} U(t,x) =0,\ t \geq 0 .
\ee
This boundary condition---together with some sort
of decay of $\fn$ at infinity---is used in all mathematical investigations
of the Vlasov-Poisson system and related systems that we are aware of,
with the exception of systems on bounded domains, considered in the
plasma physics case.

In the context of general relativity it is traditional to consider also
cosmological solutions. An isolated system can be thought of as a
more or less localized deviation from flat space, whereas a cosmological
solution is a deviation from a homogeneous state, usually one with
non-zero, spatially constant mass density. The present
paper investigates this situation in a nonrelativistic, Newtonian
setting. Our main result is global existence and uniqueness of classical
solutions for initial data which represent spatially periodic deviations
from the homogeneous state.
For the applications in cosmology see \cite{Pe}.
Note that very little is known about the existence of Newtonian
cosmological models without symmetry. The only available result
is a local in time existence theorem for the Euler-Poisson system
\cite{Br}.
The present paper proceeds as follows:
In the next section we briefly describe spatially homogeneous solutions of
the Vlasov-Poisson system where $f$ is locally isotropic in $v$. There
are essentially two types of such solutions: solutions which exist
for all future times and are expanding in a certain sense, and
solutions which first expand, then contract, and collapse to a singular
state in finite time. In general relativity, these would correspond
to an ``open'' or ``closed'' universe respectively. Next we derive
the Vlasov-Poisson-type system which holds for a deviation from such
a homogeneous state. In Sect.~3 we fix some notation. A local
existence and uniqueness result for classical solutions
to the initial value problem is
established in Sect.~4, together with a continuation criterion
for such solutions. An important ingredient in the proof of global
existence is a bound on the kinetic energy of the system. Such a bound
will be established in Sect.~5. In the last section we finally prove
that the solutions exist globally in time. In the case of an open
universe, where the homogeneous background state exists for
all future times, the meaning of this statement is clear, and, in
addition, we obtain information on the asymptotic growth of
the deviation. In the case of a closed universe, where the
homogeneous state exists only for a finite time, ``global''
means that the deviation exists as long as the homogeneous
state. This should not be mistaken for a merely local
existence result: it might have been thought that the deviation develops
a singularity before the final ``big crunch'' of the homogeneous
state, but our result shows that this is not so.

To conclude this introduction we give some references to the
literature where we restrict ourselves to investigations
of the existence of global, classical solutions.
In the case of an isolated system, global existence
of classical solutions for the Vlasov-Poisson system with general
initial data was first obtained by Pfaffelmoser \cite{Pf}.
His proof was greatly simplified and his asymptotic estimates
improved by Schaeffer \cite{Sch1} and Horst \cite{Ho2}, see
also \cite{BR2} for a spatially periodic version of these results
in the plasma physics case.
With completely different methods, Lions and Perthame \cite{LP}
also obtained global existence for the Vlasov-Poisson system.
A version of this system incorporating special relativistic effects
was considered
by Glassey and Schaeffer who showed that solutions of
the relativistic Vlasov-Poisson system  develop singularities
in finite time in the stellar dynamic case
if the initial data have negative energy \cite{GlSch}.
However, this system has the defect that it is neither Galilei nor
Lorentz invariant. A fully Lorentz invariant system, considered
in the plasma physics case, is the relativistic Vlasov-Maxwell
system, where the question of global existence of classical solutions
for general data is open. We refer to \cite{R} and the references
therein for global results for special data. A fully relativistic
version with gravitational interaction, the Vlasov-Einstein system,
was investigated only recently in \cite{RR}, and global existence
was obtained for small, spherically symmetric data. As mentioned
earlier, all these results treat the case of an isolated system.
 %
 %
\section{Homogeneous solutions and the system for deviations from these}
\setcounter{equation}{0}

In this section we first investigate a class of solutions to
the Vlasov-Poisson system (\ref{vl}), (\ref{po}), ({\ref{rho})
which are homogeneous in $x$ and locally isotropic in $v$.
Then we shall derive a system of equations which is satisfied
by deviations from such a homogeneous state.

For a nonnegative function $H \in C^1_c (\R)$ we set
\be \label{fndef}
f_0 (t,x,v) :=
H\left( a^2(t) \Bigl|v - \frac{\dot a (t)}{a(t)} x\Bigr|^2 \right)
\ee
where $a$ is a positive, scalar function to be determined later. We obtain
\[
\rho_0(t,x) = \int f_0(t,x,v)\, dv = a^{-3}(t) \int H(v^2)\, dv ,
\]
and, after normalizing
\[
\int H(v^2)\, dv =1,
\]
we have the homogeneous mass density
\be \label{rndef}
\rho _0(t) = a^{-3}(t),\ t\geq 0.
\ee
A solution of the corresponding Poisson equation is then given by
\be \label{undef}
U_0(t,x) := \frac{2\pi}{3} a^{-3}(t)\, x^2,\ t\geq 0,\ x\in \R^3.
\ee
This potential does not satisfy the boundary condition (\ref{bc}),
but this cannot be expected from a solution of the Poisson equation
(\ref{po}) with a spatially constant, non-zero density.
We note that
\[
\dx U_0(t,x) = \frac{4 \pi}{3} a^{-3}(t)\, x,
\]
and it remains to determine the function $a$ in such a way
that $f_0$ satisfies the Vlasov equation with force term
$-\dx U_0$. This is the case if the quantity
\[
a^2(t) \Bigl|v - \frac{\dot a (t)}{a(t)} x\Bigr|^2
\]
satisfies this equation. A short computation shows that
\beas
&&\Bigl( \dt + v\cdot \dx - \dx U_0 \cdot \dv\Bigr)
a^2(t) \Bigl|v - \frac{\dot a (t)}{a(t)} x\Bigr|^2 =\\
&& \qquad \qquad \qquad
- 2 a(t) \left( v - \frac{\dot a(t)}{a(t)} x\right)\cdot x
\left( \ddot a(t) + \frac{4 \pi}{3} a^{-2}(t) \right)
\eeas
which is zero for $t\geq 0,\ x,v \in \R^3$ if and only if
$a$ is a solution of the differential equation
\be \label{agl}
\ddot a + \frac{4 \pi}{3} a^{-2} =0 .
\ee
This equation describes radial motion in the gravitational field of a
point mass, and it is well known and obvious that the quantity
\[
E_a := \frac{1}{2} \dot a(t)^2 - \frac{4 \pi}{3} a^{-1} (t)
\]
is conserved along solutions.
Let $a:[0,T_a[ \to ]0,\infty[$
be a solution of Eqn.\ (\ref{agl}) on its right maximal
interval of existence with  $a(0)=1$. We can distinguish the following
two cases, depending on whether the energy $E_a$ is
positive or negative:

\bigskip

\noindent
{\bf Case A},
$\dot a(0) \geq \sqrt{8\pi/3}$: The solution of Eqn.\ (\ref{agl}) exists
on $[0,\infty[$ and $\dot a (t) >0$ for $t\geq 0$, i.\ e.\
the solution expands forever.

\smallskip

\noindent
{\bf Case B},
$\dot a(0) < \sqrt{8\pi/3}$: The maximal existence time $T_a$ is finite,
$\lim_{t \to T_a} a(t) =0$, and
there exists a time $t_a \in [0,T_a[$ such that
$\dot a(t) >0$ for $t\in [0,t_a[$ and
$\dot a(t) <0$ for $t\in ]t_a,T_a[$, i.\ e.\ the solution
expands up to the time $t_a$ and
then contracts and collapses to a singular state in finite time.

\bigskip

\noindent
We remark that in both cases the function $a$ can be extended
into the past only for a finite time, after which it collapses to
a singular state, the ``big bang''.

We will now formulate the system of equations which is satisfied
by perturbations of the homogeneous state $f_0,\rho_0,U_0$.
Let
\[
f=f_0 + g,\ \rho =\rho_0 + \s,\ U = U_0 + W
\]
be a solution of the Vlasov-Poisson system (\ref{vl}), (\ref{po}), (\ref{rho}).
A short computation shows that this is the case if and only if
$g,\s,W$ solves the system
\be \label{pvl}
\dt g + v\cdot \dx g -
\left(\dx W + \frac{4\pi}{3} a^{-3}(t)\, x \right)\cdot \dv g =
\dx W \cdot \dv f_0 ,
\ee
\be \label{ppo}
\lap W = 4\pi \s,
\ee
\be \label{prho}
\s (t,x):= \int g(t,x,v)\,dv .
\ee
Essentially, one can now consider two types of such deviations:
One can consider deviations from the homogeneous background which
are in some sense localized, say, the support of $g$ is compact initially,
or one can consider deviations which
extend over the whole universe in some periodic way. However, due to the
non-zero right hand side in Eqn.\ (\ref{pvl}) an initially compact support of
the perturbation $g$ is not preserved by the time evolution of the system
(\ref{pvl}), (\ref{ppo}), (\ref{prho}).
Also, it seems physically more convincing to have the deviation from
the homogeneous state extend over the whole universe instead of only
some part of it. The explicit $x$-dependence of the coefficient in
Eqn.\ (\ref{pvl}) seems to preclude the possibility of periodic
solutions to this system but this can be overcome by the
following transformation of variables:
\be \label{tra}
\begin{array}{ccl}
\tilde x & = & a^{-1} (t) \,x \\
\tilde v & = & v - a^{-1}(t) \dot a(t)\, x
\end{array}
\ \ \mbox{i.\ e.} \ \
\begin{array}{ccl}
x & = & a (t)\, \tilde x \\
v & = & \tilde v + \dot a(t)\, \tilde x
\end{array} ,
\ee
and
\be \label{trag}
\tilde g (t,\tilde x,\tilde v) = g(t,x,v) =
g(t, a(t) \tilde x,\tilde v + \dot a(t) \tilde x) ,
\ee
\be \label{tras}
\tilde \s (t,\tilde x) = \s (t,x) = \s (t, a(t) \tilde x),
\ee
\be \label{traw}
\tilde W (t,\tilde x) = W(t,x) = W(t, a(t) \tilde x).
\ee
Then $\tilde g, \tilde \s,\tilde W$ satisfy the system
\[
\dt \tilde g + \frac{1}{a} \tilde v\cdot \partial_{\tilde x} \tilde g -
\frac{1}{a} \left(\partial_{\tilde x} \tilde W +
\dot a \tilde v  \right)\cdot \partial_{\tilde v} \tilde g =
2 a H'(a^2 \tilde v^2)\, \tilde v \cdot \partial_{\tilde x} \tilde W,
\]
\[
\lap \tilde W = 4\pi a^2 \tilde \s,
\]
\[
\tilde \s (t,\tilde x):= \int \tilde g(t,\tilde x,\tilde v)\,d\tilde v .
\]
Since throughout the present paper we will be concerned with
this system in the transformed variables, we now drop the tilde
and obtain the following system, describing the time evolution
of a deviation from the homogeneous, locally isotropic state:
\be \label{pvlt}
\dt  g + \frac{1}{a}  v\cdot \partial_{ x}  g -
\frac{1}{a} \Bigl(\partial_{ x}  W +
\dot a  v  \Bigr)\cdot \partial_{ v}  g =
2 a H'(a^2  v^2)\,  v \cdot \partial_{ x}  W,
\ee
\be \label{ppot}
\lap  W = 4\pi a^2  \s,
\ee
\be \label{prhot}
 \s (t, x):= \int  g(t, x, v)\,dv ,
\ee
where it should be kept in mind that $t\geq 0, x,v \in \R^3$ are related
to the original variables by the transformation (\ref{tra}).
We shall be interested in solutions of this system which are
periodic in $x$; note that the coefficients in Eqn.\ (\ref{pvlt})
no longer depend on $x$ explicitly. It should be noted that
the resulting solutions
to the original Vlasov-Poisson system are not spatially periodic,
since we have rescaled
the space variable in a time-dependent manner; the change in the
velocity variable simply means that the bulk velocity
$ a^{-1}(t) \dot a(t) x$ of the homogeneous background solution
is shifted into the origin of the velocity space.

To conclude this section, we transform the original Vlasov-Poisson
system (\ref{vl}), (\ref{po}), (\ref{rho}) according to
(\ref{tra}), drop the tilde and obtain:
\be \label{vlt}
\dt f + \frac{1}{a} v\cdot \dx f -
\frac{1}{a} \left(\dx U - \frac{4 \pi}{3} a^{-1} x +
\dot a v \right)\cdot \dv f =0,\
\ee
\be \label{pot}
\lap U = 4\pi a^2 \rho,
\ee
\be \label{rhot}
\rho (t,x):= \int f(t,x,v)\,dv .
\ee
Throughout the paper we make the following assumption:
\[
\mbox (A)\
\left\{
\begin{array}{l}
a:[0,T_a[ \to ]0,\infty[\ \mbox{is a maximal solution of Eqn.\ (\ref{agl})
with}\ a(0)=1 \\
H \in C^1_c (\R),\ H\geq 0,\ \int H(v^2)\, dv =1,\ \mbox{and}\
H(v^2)=0,\ \n{v} \geq u_0 >0
\end{array}
\right.
\]
We define $f_0,\rho_0,U_0$ by
\be \label{homdef}
f_0(t,x,v):= H(a^2(t) v^2),\ \rho_0(t):= a^{-3}(t),\ U_0(t,x):=
\frac{2\pi}{3} a^{-1}(t) x^2,
\ee
which are the equations (\ref{fndef}), (\ref{rndef}), (\ref{undef}),
transformed according to (\ref{tra}).
Then $f_0,\rho_0,U_0$ is a solution of the transformed Vlasov-Poisson
system (\ref{vlt}), (\ref{pot}), (\ref{rhot}), which we refer
to as the homogeneous, locally isotropic background state of the
Vlasov-Poisson system.
If $g, \s,W$ satisfy the system (\ref{pvlt}), (\ref{ppot}), (\ref{prhot})
then $f=f_0 + g$, $\rho = \rho_0 + \s$, $U = U_0 + W$ is a solution
of  (\ref{vlt}), (\ref{pot}), (\ref{rhot}), and, via the transformation
(\ref{tra}), also a solution of the original Vlasov-Poisson system
(\ref{vl}), (\ref{po}), (\ref{rho}), which we refer to as
a cosmological solution of the Vlasov-Poisson system, and which no longer
has to be homogeneous or isotropic.
 %
 %
\section{Some notation}
\setcounter{equation}{0}

Let $Q:= [0,1]^3$ and $S:= Q \times \R^3$. We need to define some
spaces of periodic functions:
\beas
{\cal P} (Q)&:=& \Bigl\{ h: \R^3 \to \R \mid h(x + \alpha ) = h(x),\
x\in \R^3,\ \alpha \in \Z^3 \Bigr\} ,\\
{\cal P} (S)&:=& \Bigl\{ h: \R^6 \to \R \mid h(x + \alpha ,v ) = h(x,v),\
x, v\in \R^3,\ \alpha \in \Z^3 \Bigr\}  , \\
C^n_\pi (Q) &:=& C^n (Q) \cap {\cal P }(Q), \\
C^n_\pi (S) &:=& C^n (S) \cap {\cal P }(S), \\
C^n_{\pi,c} (S) &:=& \Bigl\{h \in C^n_\pi (S) \mid
\exists u\geq 0 : h(x,v)=0,\ \n{v} >u \Bigr\} .
\eeas
For $p\in [1,\infty]$ we denote by $\nn{\cdot}_p $ the usual $L^p$-norm,
where the integral (or supremum) extends over $Q$ or $S$ as the case may be,
and
\beas
L^p_\pi (Q) &:=& L^p (Q) \cap {\cal P} (Q) ,\\
L^p_\pi (S) &:=& L^p (S) \cap {\cal P} (S) ,
\eeas
where we identify functions in ${\cal P} (Q)$ and ${\cal P} (S)$
with their restrictions to $Q$ or $S$ respectively.
For the pair $(x,v)$ we frequently write $z$.
 %
 %
\section{Local existence and continuation of solutions}
\setcounter{equation}{0}

In this section we shall prove the existence of local
solutions of the system (\ref{pvlt}), (\ref{ppot}), (\ref{prhot})
for initial data in $C^1_{\pi,c}(S)$. To do this, we have to be
able to solve the equations (\ref{pvlt}) and (\ref{ppot})
seperately. We start by investigating the Poisson equation for
a given, spatially periodic density.

\begin{lemma} \label{green}
There exists an even function
$G\in C^\infty (\R^3 \setminus \Z^3) \cap {\cal P}(Q)$ with the following
properties:
\begin{itemize}
\item[(a)]
For $\sigma \in C^1_\pi (Q)$ with $\int_Q \s (x)\, dx =0$ the
function
\[
W(x):= \int_Q G(x-y) \s (y)\, dx = \int_{x'+Q} G(x-y) \s (y) \, dy,\
x, x'\in \R^3
\]
is in $C^2_\pi (Q)$ the unique solution of
\[
\lap W = 4 \pi \s
\]
with $\int_Q W(x)\, dx =0$.
\item[(b)]
There exists a function
$G_0 \in  C^\infty ((\R^3 \setminus \Z^3)\cup\{(0,0,0)\})$ such that
\[
G(x) = - \frac{1}{\n{x}} + G_0 (x),\ x \in \R^3 \setminus \Z^3 .
\]
\end{itemize}
\end{lemma}
For a proof of this result we refer to \cite{BR1}.
Next we note some estimates for the potential $W$ which can be
proved by treating the singular term in the Green's function $G$
as in \cite[Props.\ 1,\ 2]{B}.

\begin{lemma} \label{west}
There exists a constant $C>0$ such that for all $\s$ as in Lemma \ref{green}
the corresponding potential $W$ satisfies the following estimates:
\begin{itemize}
\item[(a)]
$\displaystyle
\nn{W}_\infty \leq C \left( \nn{\s}_1^{2/3} \nn{\s}_\infty^{1/3}
+ \nn{\s}_1 \right) $
\item[(b)]
$\displaystyle
\nn{\dx W}_\infty \leq C \left( \nn{\s}_1^{1/3} \nn{\s}_\infty^{2/3}
+ \nn{\s}_1 \right) $
\item[(c)]
$\displaystyle
\nn{\dx ^2 W}_\infty \leq C \left( (1+R^{-3}) \nn{\s}_1 +
(1 + \ln R/d ) \nn{\s}_\infty + d\, \nn{\dx \s}_\infty \right)$
for any $d,R$ with  $0 < d\leq R \leq 1/2 $
\item[(d)]
$\displaystyle
\nn{\dx ^2 W}_\infty \leq C \bigl( \nn{\s}_1 +
(1 + \nn{\s}_\infty)
(1 + \ln ^\ast \nn{\dx \s}_\infty/2 ) \bigr)$, \newline
where
\[
\ln^\ast s := \left\{ \begin{array}{ccc}
s &,& s \leq 1\\
1 + \ln s &,& s > 1
\end{array} \right. ,
\]
\item[(e)]
$\displaystyle
\nn{\dx W}_\infty \leq C \left( \nn{\s}_{5/3}^{5/9}\, \nn{\s}_\infty^{4/9}
+ \nn{\s}_1 \right) $ .
\end{itemize}
\end{lemma}
Next we investigate Eqn.\ (\ref{pvlt}) for a given potential $W$.
To this end, we first consider the corresponding characteristic
system
\be \label{cha}
\dot x = \frac{1}{a(t)} v,\
\dot v = - \frac{1}{a(t)} \Bigl( \dx W (t,x) + \dot a(t) v \Bigr) .
\ee

\begin{lemma} \label{chasol}
Let $0< T \leq T_a$ and $W \in C^{0,2}_\pi ([0,T[ \times Q)$. Then the
following holds:
\begin{itemize}
\item[(a)]
For $z = (x,v) \in \R^6$ and $t\in [0,T[$ there exists a unique solution
\[
[0,T [ \ni s \mapsto \bigl(X(s,t,x,v),V(s,t,x,v)\bigr) = Z(s,t,z)
\]
of the characteristic system (\ref{cha}) with
$Z(t,t,z)=z$ .
\item[(b)]
$Z \in C^1 ([0,T[^2 \times \R^6)$, for every $s,t \in [0,T[$ the mapping
$Z(s,t,\cdot)$ is a $C^1$-diffeomorphism of $\R^6$, and
\[
Z(s,t,x +\alpha,v) = Z(s,t,x,v) + (\alpha,0),\
\alpha \in \Z^3 .
\]
\item[(c)]
$\displaystyle
\det \partial_z Z(s,t,z) = \left(\frac{a(t)}{a(s)}\right)^3,\ s,t \in [0,T[.$
\end{itemize}
\end{lemma}

\prf
Part (a) and the first two assertions in (b) are standard, the third assertion
in (b) follows by uniqueness and the periodicity of $W$.
For (c), note that
\[
\div_z \left( \frac{1}{a} v, - \frac{1}{a} (\dx W + \dot a v ) \right) =
-3 \frac{\dot a}{a}\; .
\]
Thus
\[
\frac{d}{ds} \bigl(\det \partial_z Z(s,t,z) \bigr)
= -3 \frac{\dot a(s)}{a(s)}\det \partial_z Z(s,t,z),
\]
and the result follows.
\prfe

\begin{lemma} \label{vlasol}
Let $W$ satisfy the assumption in Lemma \ref{chasol} and take
$\gn \in C^1_{\pi,c} (S)$ such that
$\gn (z) + H(v^2) \geq 0$ for $z \in \R^6$.
Then
\bea
g(t,z)&:=& \gn (Z(0,t,z)) +
2 \int_0^t a(s) \Bigl(H'(a^2 v^2)\, v \cdot \dx W\Bigr)(s,Z(s,t,z)) ds
\label{gdef} \\
&=&
\gn (Z(0,t,z)) + H(V^2 (0,t,z)) - H(a^2(t) v^2),\ t\in [0,T[,\ z\in \R^6,
\ \ \ \label{gdefs}
\eea
defines the unique solution of Eqn.\ (\ref{pvlt}) in $C^1([0,T[\times \R^6)$
with $g(0)=\gn$. The solution $g$ has the following properties:
\begin{itemize}
\item[(a)]
$\displaystyle
g \in C([0,T[; C^1_{\pi,c} (S)),$
\item[(b)]
$ \int_S \gn (z) \, dz =0$ implies $\int_S g(t,z)\, dz =0,\ t\in [0,T[$,
\item[(c)]
$ \displaystyle
\nn{g(t)}_\infty \leq \nn{\gn}_\infty + 2 \nn{H}_\infty,\ t\in [0,T[$,
\item[(d)]
$ \displaystyle
\nn{g(t)}_1 \leq (\nn{\gn}_1 + 2) \, a^{-3} (t),\ t\in [0,T[$.
\end{itemize}
\end{lemma}

\prf
The fact that Eqn.\ (\ref{gdef}) defines the unique $C^1$-solution of
Eqn.\ (\ref{pvlt}) with $g(0)=\gn$ is standard if $H \in C^2 (\R)$. Since
$- H(a^2(s) V^2(s,t,z))$ is a primitive of the integrand in (\ref{gdef}),
Eqn.\ (\ref{gdefs}) defines the same function, and an approximation
argument for $H$ shows that it is the unique $C^1$-solution also if
$H \in C^1 (\R)$. The assertion in (a) follows from the periodicity
of $\gn$ and Lemma \ref{chasol}. To see the assertion in (b), integrate
Eqn.\ (\ref{pvlt}) with respect to $z$, observe that the right hand side
in that equation is odd in $v$, and integrate by parts with respect
to $x$ and $v$ respectively to obtain the equation
\[
\frac{d}{dt} \int_S g(t,z)\, dz =
- 3 \frac{\dot a(t)}{a(t)} \int_S g(t,z)\, dz,\ t \in [0,T[ ;
\]
observe that no boundary terms occur due to the compact support in $v$ and the
periodicity in $x$ respectively. The assertion in (b) follows but
observe that we obtain no information on $\nn{g(t)}_1$ from
this since $g$ is not positive.
The assertion in (c) is obvious by (\ref{gdefs}). To prove (d),
define $f:= f_0 + g$. Then $f$ satisfies the equation
\[
\dt f + \frac{1}{a} v \cdot \dx f - \frac{1}{a} ( \dx W + \dot a v)
\cdot \dv f =0
\]
so that
\[
f(t,z)= f(0,Z(0,t,z)) = H(V^2(0,t,z)) + \gn (Z(0,t,z)) \geq 0
\]
by assumption on $\gn$, and $f(t) \in C^1_\pi (S),\ t\ \in [0,T[$.
The same argument as above proves that
\[
\frac{d}{dt} \int_S f(t,z)\, dz =
- 3 \frac{\dot a(t)}{a(t)} \int_S f(t,z)\, dz,\ t \in [0,T[ .
\]
Thus
\[
\nn{f(t)}_1 = \nn{f(0)}_1 a^{-3} (t),
\]
and since
\[
\nn{f_0(t)}_1 = a^{-3}(t),
\]
the assertion in (d) follows.
\prfe
We are now ready to prove local existence of classical solutions
to the initial value problem for the system (\ref{pvlt}), (\ref{ppot}),
(\ref{prhot}).

\begin{thm} \label{locex}
Let $\gn \in C^1_{\pi,c} (S)$ with $\int_S \gn (z)\, dz =0$ and
$\gn (x,v) + H(v^2) \geq 0,\ (x,v) \in \R^6$.
Then there exists a unique maximal solution
\[
g \in C^1 ([0,T[ \times \R^6) \cap C([0,T[; C^1_{\pi,c} (S))
\]
of the system  (\ref{pvlt}), (\ref{ppot}), (\ref{prhot})
with $g(0)= \gn$, $0 < T \leq T_a$. Moreover, if
\[
\sup \{ \n{v} \mid (x,v) \in \supp g(t),\ 0 \leq t < T \} < \infty
\]
then $T = T_a$, i.\ e.\ the solution exists as long as the
homogeneous state.
\end{thm}

\prf
The method of proof is quite standard, and we only indicate the main steps;
for an isolated system the corresponding arguments can be found in \cite{B}.

Consider the following iterative scheme: Define
\[
g_0 (t,z) := \gn (z),\ t\in [0,T_a[,\ z\in \R^6,
\]
and if
\[
g_n \in C^1 ([0,T[ \times \R^6) \cap C([0,T[; C^1_{\pi,c} (S))
\]
is already defined,
\[
\s_n(t,x) := \int g_n (t,x,v)\, dv,
\]
\[
W_n (t,x) := a^2(t) \int_Q G(x-y) \s_n(t,y) \, dy
\]
for $t \in [0,T_a[,\ x \in \R^3$, let $Z_{n+1}$ be defined according
to Lemma \ref{chasol} with $W_n$ instead of $W$, and
\[
g_{n+1}(t,z):=
\gn (Z_{n+1}(0,t,z)) + H(V^2_{n+1} (0,t,z)) - H(a^2(t) v^2),
\ t\in [0,T_a[,\ z\in \R^6.
\]
It follows from the above Lemmata that this scheme is well defined.
With
\beas
P_n (t)&:=& \sup \left\{ \n{V_{n-1} (s,0,z)} \mid
z \in \supp \gn \cup \R^3 \times B_{u_0}(0),\,
0\leq s \leq t \right\}\\
&&  + u_0 a^{-1}(t),\ t\in [0,T_a[,
\eeas
it follows that
\[
g_n (t,x,v) =0,\ \n{v} > P_n (t) .
\]
Using the characteristic system and the estimates from Lemma \ref{west}
and Lemma \ref{vlasol} it can be shown by induction that
\[
P_n (t) \leq P(t),\ t\in [0,\delta[,\ n\in \N
\]
where $P:[0,\delta [ \to \R^+$ is the right maximal solution of the equation
\be \label{pgl}
P(t) = P_0 + C(\gn) \int_0^t \left(P^2 (s) +
\frac{\n{\dot a (s)}}{a(s)} P(s) + a^{-2}(s)\right) ds + u_0 a^{-1}(t),
\ee
$\delta \leq T_a$.
Here $C(\gn)$ depends only on $\nn{\gn}_1$ and $\nn{\gn}_\infty$, and
\[
P_0 := \sup \{ \n{v} \mid z \in \supp \gn \cup \R^3 \times B_{u_0}\}.
\]
On the interval $[0,\delta[$ the iterative scheme can now be seen to converge.
First observe that
\[
\nn{\s_n(t)}_\infty \leq C P^3 (t),\
\nn{\dx W (t)}_\infty \leq C \left(a^{-1}(t) + a(t) P^2 (t)\right) ,
\ n\in \N,\ t\in [0,\delta[.
\]
Next it can be shown that on any compact subinterval $[0,\delta_0]$ of
$[0,\delta[$
\[
\nn{\dx \s_n (t)}_\infty \leq C,\ \nn{\dx^2 W (t)}_\infty \leq C,
\ n\in \N.
\]
This can be seen from the following Gronwall-type inequality
\[
\nn{\dx \s_n (t)}_\infty \leq C \exp \left( \int_0^t \ln^\ast
\frac{\nn{\dx \s_{n-1} (s)}_\infty}{2} ds \right),
\]
which follows from Lemma \ref{west} (d).
Now define
\[
\alpha_n (t):= \sup \bigl\{\n{X_{n+1} - X_n}(s,t,z)+ \n{V_{n+1} - V_n}(s,t,z)
\mid 0 \leq s \leq t,\ z \in \R^6 \bigr\}
\]
for $t \in [0,\delta_0]$.
Using the uniform bounds which hold on the interval $[0,\delta_0]$
another Gronwall argument shows that
\[
\alpha_n(t) \leq C \int_0^t \alpha_{n-1} (s) \, ds ,\ 0 \leq t\leq \delta_0,
\ n \geq 1.
\]
Thus, the sequence
$Z_n$ is a uniform Cauchy sequence on $[0,\delta_0]^2 \times \R^6$.
The convergence of $g_n(t)$, $\s_n(t)$, $W_n(t)$
in $C_\pi (S)$, $C_\pi (Q)$, and $C^1_\pi (Q)$ respectively follows,
uniformly on $[0,\delta_0]$. Using Lemma \ref{west} (c) it can be seen
that  $W_n(t)$ is actually a Cauchy sequence in  $C^2_\pi (Q)$,
uniformly on $[0,\delta_0]$. Thus, the function $g$
is $C^1$, and since $\delta_0 < \delta$ is arbitrary
we obtain the desired classical solution on the interval $[0,\delta[$.

To obtain uniqueness, the Gronwall arguments indicated above
may be used on the difference of two solutions with
the same initial datum.

Let $g \in C^1([0,T[ \times \R^6)$ be the unique maximal solution,
and assume that $T < T_a$ and
\[
P^\ast := \sup \{ \n{v} \mid (x,v) \in \supp g(t),\ 0 \leq t < T \} < \infty.
\]
Then by Lemma \ref{vlasol}
\[
\nn{g(t)}_1 + \nn{g(t)}_\infty \leq C,\ t\in [0,T[.
\]
Consider now Eqn.\ (\ref{pgl}), but with $g(t_0)$ instead of
$\gn$ and $P^\ast$ instead of $P_0$. Then the function $P$
exists on an interval $[t_0, t_0 + \delta[$, the length of which
is bounded from below by a positive constant,
which is independent of $t_0 \in [0,T[$. Moving $t_0$ close enough
to $T$ the above proof shows that there exists a solution
on the interval $[t_0,t_0 + \delta[$ which extends the
solution $g$ beyond the time $T$. This is a contradiction,
and the proof is complete.
\prfe

 %
 %
\section{An energy estimate}
\setcounter{equation}{0}

Throughout the rest of this paper $(g,\s,W)$ denotes
a solution of the system (\ref{pvlt}), (\ref{ppot}), (\ref{prhot})
on a maximal interval of existence $[0,T[ \subset [0,T_a[$
as obtained in Theorem \ref{locex}. By $C$ we shall denote
constants which may change from line to line and may depend on $\gn$
but never on $t$.
We want to establish a bound on the kinetic energy which in turn
shall give a bound on $\nn{g(t)}_{5/3}$. This is desirable because it allows
the force term $\dx W$ to be bounded by a lower power of $\nn{\s (t)}_\infty$,
cf.\ Lemma \ref{west} (e). However, the latter argument immediately
runs into  the problem that $g$ is not necessarily positive, and
a bound on the kinetic energy of $g$ is useless when trying
to control $\nn{g(t)}_{5/3}$. To avoid this problem, we consider
$f:= f_0 + g$ instead. This function
satisfies the equation
\[
\dt f + \frac{1}{a} v \cdot \dx f - \frac{1}{a}
(\dx W + \dot a v) \cdot \dv f = 0.
\]
Thus it is constant along characteristics, and since it is nonnegative
initially by assumption on $\gn$, is nonnegative for all $t\in [0,T[$.
Now define
\[
E_{kin} (t):= \int_S v^2 f(t,z)\, dz,
\]
\[
E_{pot} (t) := \int_Q W(t,x) \s (t,x)\, dx = - \frac{1}{4 \pi a^2 (t)}
\int_Q \n{\dx W(t,x)}^2 dx ,\ t \in [0,T[.
\]
If we differentiate the kinetic energy with respect to time,
use the differential equation for $f$, and
integrate by parts we obtain
\be \label{eee}
\frac{d}{dt} E_{kin} (t) = - 5 \frac{\dot a (t)}{a(t)} E_{kin} (t)
+ \frac{2}{a(t)} \int_Q W (t,x) \div j(t,x)\,dx,
\ee
where
\[
j(t,x):= \int v g(t,x,v)\, dv ;
\]
$f_0$, being even in $v$, does not contribute to the mass current density.
Integrating (\ref{pvlt}) with respect to $v$ yields
\[
\dt \s + \frac{1}{a} \div j + 3 \frac{\dot a}{a} \s =0 ,
\]
and inserting this into (\ref{eee}) we obtain
\be \label{kine}
\frac{d}{dt} E_{kin} (t) = - 5 \frac{\dot a (t)}{a(t)} E_{kin} (t)
- 2 \int_Q W (t,x) \dt \s (t,x)\,dx - 6 \frac{\dot a}{a} E_{pot}(t) .
\ee
Differentiating the potential energy yields
\be \label{pote}
\frac{d}{dt} E_{pot} (t) = 2 \int_Q W (t,x) \dt \s (t,x)\,dx
+ 2 \frac{\dot a}{a} E_{pot}(t) .
\ee
{}From Eqns.\ (\ref{kine}) and (\ref{pote}) it follows that
\be \label{econ}
a^5 (t) E(t) =
E(0) + \int_0^t \dot a(s) a^4 (s) E_{pot} (s) ds ,\ t \in [0,T[,
\ee
where $E := E_{kin} + E_{pot}$ denotes the total energy.
To exploit Eqn.\ (\ref{econ}) we need to estimate the potential energy
in terms of the kinetic energy. This argument is due to Horst
\cite[(5.7)]{Ho1}, and we only write down the result for our
present situation:

\begin{lemma} \label{potkin}
For $t \in [0,T[$,
\[
\n{E_{pot} (t) } \leq C \left( a^{-4} (t) + a^{-3/2} (t) E_{kin} (t)^{1/2}
\right) .
\]
\end{lemma}
To continue, it is convenient to distinguish between Case A and Case B.
In Case A, $\dot a >0$, and since the potential energy is negative,
Eqn.\ (\ref{econ}) implies that
\[
E(t) \leq a^{-5}(t) E(0).
\]
With Lemma \ref{potkin} this yields
\[
E_{kin} (t) \leq C \left( a^{-4} (t) + a^{-3/2} (t) E^{1/2}_{kin} (t) \right),
\ t\in [0,T[,
\]
which gives a bound on the kinetic energy in Case A.
In Case B where $\dot a$  eventually becomes negative, we have to argue
differently. Observing that in this case $a$ is bounded and
$T \leq T_a < \infty$ we obtain from (\ref{econ}) and
Lemma \ref{potkin} the estimate
\beas
a^5(t) E_{kin} (t) &\leq& C \biggl( 1 + \int_0^t
\n{\dot a (s)} a^4(s) \left(a^{-4}(s) + a^{-3/2} E_{kin}(s))^{1/2}\right) ds\\
&& \qquad +  a(t) + a^{7/2}(t) E_{kin}(t))^{1/2} \biggr)\\
&\leq&
C \left( 1 + \max_{0 \leq s \leq t} \left( a^5 (s) E_{kin} (s) \right)^{1/2}
\right) \left( 1 + \int_0^t \n{\dot a (s)} ds \right) \\
&\leq&
C \left( 1 + \max_{0 \leq s \leq t} \left( a^5 (s) E_{kin} (s) \right)^{1/2}
\right)
\eeas
which yields a bound for the kinetic energy in Case B.
Thus we have proved the following result:

\begin{lemma} \label{kinab}
For $t \in [0,T[$,
$E_{kin} (t) \leq C a^{-3} (t)$ in Case A, and
$E_{kin} (t) \leq C a^{-5} (t)$ in Case B.
\end{lemma}
By an interpolation argument and estimating $g$ in terms of $f + f_0$,
these estimates yield the following bounds on $\nn{\s (t)}_{5/3}$:

\begin{cor} \label{lpbound}
For $t \in [0,T[$,
$\nn{\s (t)}_{5/3}  \leq C a^{-9/5} (t)$ in Case A, and
$\nn{\s (t)}_{5/3}  \leq C a^{-3} (t)$ in Case B.
\end{cor}
Thus we see that although we have no energy conservation for
our problem, the usual bounds on the kinetic energy
and the consequences thereof still hold.

 %
 %
\section{Global existence}
\setcounter{equation}{0}

In this section we prove the following global existence result
which is the main result of the present paper:

\begin{thm} \label{glex}
Let $\gn \in C^1_{c,\pi} (S)$ with $\int_S \gn (z) dz =0$ and
$\gn (z) + H(v^2) \geq 0$ for $z\in \R^6$. Then there exists
a unique solution $g \in C^1 ([0,T_a[ \times \R^6) \cap C([0,T_a[;
C^1_\pi (S))$ of the system (\ref{pvlt}), (\ref{ppot}), (\ref{prhot})
with $g(0)= \gn$. Moreover, in Case A, where $T_a=\infty$,
we have the estimate
\[
P(t) =O (t^{2+\delta}), t\geq 0,
\]
for any $\delta >0$ where
\[
P(t):= \sup \left\{ \n{v} \mid
z \in \supp g (s),\  0 \leq s \leq t \right\} .
\]
\end{thm}

\prf
The proof makes use of the ideas of Schaeffer \cite{Sch1} and some unpublished
simplifications of these \cite{Sch2}. Similar estimates have also been
used in \cite{Ho2}.
We first consider Case A. Along any characteristic we have the equation
\be \label{av}
\frac{d}{ds} \left( a(s) v(s) \right) = \dot a(s) v(s) - \dx W (s,x(s))
- \dot a(s) v(s) = - \dx W(s,x(s)) .
\ee
By Lemma \ref{west} (e) and Cor.~\ref{lpbound}
\be \label{dxw}
\nn{\dx W(s)}_\infty \leq C a^2(s) \left( a^{-1}(s) \nn{\s (s)}_\infty ^{4/9}
+ \nn{\s (s)}_1 \right)
\leq C_1 a(s) P(s)^{4/3},
\ee
and thus
\be \label{avab}
\n{a(s_1) v(s_1) - a(s_2) v(s_2)} \leq C_1 a(t) P(t)^{4/3} \n{s_1 -s_2},\
0 \leq s_1,s_2 \leq t .
\ee
Let $(X,V)$ denote a fixed characteristic which hits the support of $g$,
and take $t \in [0,T[$ and $\Delta \in [0,t]$, $T$ being the
length of the maximal existence interval of the local solution
$g$ corresponding to the initial value $\gn$. Then
\beas
&&\n{a(t) V(t) - a(t-\Delta) V(t-\Delta)}\\
&&\leq
\int_{t-\Delta}^t a^2(s) \biggl( \int_{X(s) + \hat Q} \int_{\R^3}
\n{g(s,x,v)} dv \frac{dx}{\n{x- X(s)}^2} \\
&& \qquad \qquad \qquad + \nn{\dx G_0}_\infty
\int_{X(s) + \hat Q} \int_{\R^3} \n{g(s,x,v)} dv\, dx \biggr) ds \\
&&\leq
C \int_{t-\Delta}^t a^{-1}(s) ds + C \int_{t-\Delta}^t  a^2 (s)
\int_{X(s) + \hat Q} \int_{\R^3}
\n{g(s,x,v)} dv \frac{dx}{\n{x- X(s)}^2} ds
\eeas
where $\hat Q := [-1/2,1/2]^3$. We split the domain of integration
of the second term into the following parts:
\beas
M_1&:=& \left\{ (s,x,v) \mid D(s,v) \leq p \right\} ,\\
M_2&:=& \left\{ (s,x,v) \mid D(s,v) > p \wedge \n{x- X(s)} \leq r D(s,v)^{-3}
\right\} ,\\
M_3&:=& \left\{ (s,x,v) \mid D(s,v) > p \wedge \n{x- X(s)} > r D(s,v)^{-3}
\right\} .
\eeas
Here $D(s,v):= \min \{ \n{v}, \n{v - V(s)} \}$, and $r>0,\ p>0$ are parameters
to be chosen later, $p \leq P(t)$. Let $I_i$ denote the contribution of the set
$M_i$ to the above integral. The integral $I_1$
is easily estimated:
\[
I_1 \leq \int_{t-\Delta}^t a^2(s) \int_{X(s) + \hat Q} \tilde \s (s,x)
\frac{dx}{\n{x- X(s)}^2} ds
\]
where
\[
\tilde \s (s,x) := \int_{D(s,v) \leq p} \n{g(s,x,v)} dv
\]
Now
\[
\nn{\tilde \s (s)}_{5/3} \leq C a^{-9/5} (s),\
\nn{\tilde \s (s)}_\infty \leq C p^3
\]
so that by Lemma~\ref{west} (e)
\[
I_1 \leq C p^{4/3} a(t) \Delta .
\]
The estimate $I_2$ is straightforward too:
\beas
I_2 &\leq&  \int_{t-\Delta}^t  a^2(s)
\int_{p \leq D(s,v) \leq 2 P(s)}\int_{\n{x-X(s)} \leq r D(s,v)^{-3}}
\n{g(s,x,v)} \frac{dx\, dv\, ds}{\n{x- X(s)}^2}  \\
&\leq&  \int_{t-\Delta}^t  a^2(s)
\int_{p \leq D(s,v) \leq 2 P(s)} r D(s,v)^{-3} dv\, ds \\
&\leq & C a^2(t) r \ln \frac{2 P(t)}{p} \Delta .
\eeas
To estimate $I_3$ we want to use the transformation of variables
\[
z'= Z(t,s,z)
\]
but since $g$ is not constant along characteristics we switch to
$f=f_0 + g$ first and obtain the estimate
\beas
I_3 &\leq&
\int_{t-\Delta}^t  a^2 (s)
\int_{X(s) + \hat Q} \int_{\R^3}
\n{f(s,x,v)} 1_{M_3} (s,z) dv \frac{dx}{\n{x- X(s)}^2} ds\\
&& +
\int_{t-\Delta}^t  a^2 (s)
\int_{X(s) + \hat Q} \int_{\R^3}
\n{f_0 (s,x,v)} dv \frac{dx}{\n{x- X(s)}^2} ds\\
&\leq&
J + C \int_{t-\Delta}^t a^{-1} (s) ds
\eeas
where $J$ denotes the integral over $f$ which remains to be estimated.
Let $\Delta \leq \min \{ t ,\frac{1}{2 \dot a(0)} \}$. Since $\dot a$
is positive and decreasing,
\[
\frac{1}{2}  a (s_2) \leq a (s_1) \leq a (s_2),\ t - \Delta \leq s_1 \leq s_2
\leq t .
\]
Using this estimate and Lemma~\ref{chasol} in the transformation
of variables introduced above we obtain the following estimate:
\[
J \leq C
\int_{t-\Delta}^t  a^2 (s)
\int_{Z(t,s,X(s) + \hat Q,\R^3)}
\n{f (t,z)} 1_{M_3} (s,Z(s,t,z)) \frac{dz\, ds}{\n{X(s,t,z)- X(s)}^2}  .
\]
Now observe that
\[
Z(t,s,X(s) + \hat Q,\R^3) \cap \supp f(t) \subset (X(t) + 2 \hat Q ) \times
\R^3 := M(t)
\]
if we assume that $P(t) \Delta \leq 1/4$ i.\ e.\
$\Delta \leq \min\{ t,\frac{1}{2 \dot a (0)}, \frac{1}{4 P(t)} \}$.
Defining
\[
J(z):= \{ s \in [t - \Delta, t] \mid (s,Z(s,t,z)) \in M_3 \}
\]
we obtain the estimate
\be \label{jest}
J \leq C
\int_{M(t)} f(t,z) \int_{J(z)} a^2 (s) \n{X(s,t,z) - X(s)}^{-2} ds\, dz.
\ee
To estimate the time integral we need to restrict the length of
the integration interval so that velocities do not change very much
during that interval. We have
\beas
\n{v(s_1) - v(s_2)} &\leq& \frac{1}{a(s_1)}\n{a(s_1) v(s_1) - a(s_2) v(s_2)}
+ \frac{\n{a(s_1) - a(s_2)}}{a(s_1)} \n{v(s_2)} \\
&\leq& C_1 \frac{a(t)}{a(s_1)} P(t)^{4/3} \Delta + \frac{\dot a(0)}{a(s_1)}
P(t) \Delta \\
&\leq& C_2 P(t)^{4/3} \Delta
\eeas
for $s_1,s_2 \in [t - \Delta, t]$; without loss of generality we may
assume $P(t) \geq 1$. Thus if we choose
\[
\Delta := \min \left\{ t, \frac{1}{2 \dot a(0)}, \frac{1}{4 P(t)},
\frac{p}{5 C_2 P(t)^{4/3}} \right\}
\]
the estimate
\[
\n{v(s_1) - v(s_2)} \leq \frac{p}{5}
\]
holds for  $s_1,s_2 \in [t - \Delta, t]$ and any characteristic which hits
$\supp g$. Now take $z \in M(t)$ and assume $J(z) \neq \emptyset$.
Then for $s \in [t - \Delta, t]$ and $\sigma \in J(z)$ the above estimate
and the definition of the set $M_3$ imply
\be \label{vcha}
\frac{4}{5} \n{ V(\sigma, t, z)} \leq \n{ V(s, t, z)}
\leq \frac{6}{5} \n{ V(\sigma, t, z)} ,
\ee
\be \label{vvcha}
\frac{3}{5} \n{ V(\sigma, t, z)- V(\sigma)} \leq \n{ V(s, t, z) - V(s)}
\leq \frac{7}{5} \n{ V(\sigma, t, z) - V(\s)} .
\ee
Define
\[
d(s) := X(s,t,z) - X(s),\ s \in [t -\Delta, t],
\]
choose $\bar s \in [t - \Delta , t]$ such that
$\n{d(\bar s)} = \min \{ \n{d(s)} \mid s \in [ t - \Delta, t] \}$
and define
\[
\bar d (s) := d(\bar s) + \dot d(\bar s) (s - \bar s),\ s \in [t -\Delta, t] .
\]
Since $d(\bar s) \cdot \dot d(\bar s) (s - \bar s) \geq 0$ for
$s \in [t -\Delta, t]$ we see that
\be \label{bard}
\n{ \bar d (s)} \geq \n{\dot d(\bar s)} \n{ s - \bar s},\
\ s \in [t -\Delta, t] .
\ee
On the other hand
\beas
\n{\ddot d(s) - \ddot{\bar d\;\;}(s)} &\leq&
\frac{2 \dot a (0)}{a^2 (s)} \n{V(s,t,z) - V(s)}\\
&& +
\frac{1}{a^2(s)} \bigl|\dx W (s, X(s,t,z)) - \dx W (s,X(s))\bigr| \\
&\leq& 2 C_2 \frac{P(s)^{4/3}}{a(s)},
\eeas
and thus
\[
\n{d(s) - \bar d(s)} \leq C_2 \frac{P(t)^{4/3}}{a(t-\Delta)} \n{s - \bar s}
\Delta \leq \frac{1}{5} \frac{p}{a(t - \Delta)} \n{s - \bar s}.
\]
By definition of the set $M_3$ and Eqn.\ (\ref{vvcha})
\[
p \leq \frac{5}{3} a(\bar s) \n{\dot d (\bar s)}
\]
so that
\[
\n{d(s) - \bar d(s)} \leq \frac{2}{3} \n{\dot d(\bar s)} \n{s - \bar s},\
s \in [t - \Delta , t].
\]
Together with Eqn.\ (\ref{bard}) this yields
\be \label{dest}
\n{d(s)} \geq \frac{1}{3} \n{\dot d(\bar s)} \n{s - \bar s},
\ s \in [t - \Delta, t] .
\ee
For $s \in J(z)$ the definition of the set $M_3$ and the estimates
(\ref{vcha}) and (\ref{vvcha}) imply that
\[
\n{d(s)} \geq C r \min \{ \n{v}, \n{v - V(t)} \} ^{-3}.
\]
Combining this estimate with (\ref{dest}) and
\[
\n{\dot d (\bar s)} = \frac{1}{a(\bar s)} \n{V(\bar s,t,z) - V(\bar s)}
\geq \frac{C}{a(\bar s)} \n{v - V(t)}
\]
yields
\[
\n{d(s)} \geq C \max \left \{ \frac{1}{a(t)} \n{v - V(t)}\n{s - \bar s},
\frac{r}{\n{v}^3},\frac{r}{\n{v-V(t)}^3} \right\}
\]
for $s \in [t - \Delta, t]$.
Inserting this estimate in the time integral in (\ref{jest})
and distinguishing the cases $\n{v} \leq \n{v - V(t)}$
and $\n{v} \geq \n{v - V(t)}$ finally yields the estimate
\[
\int_{J(z)} a^2(s) \n{X(s,t,z) - X(s)}^{-2} ds \leq C a^3 (t) r^{-1} \n{v}^2.
\]
Thus by Lemma \ref{kinab}
\beas
J&\leq& C r^{-1} a^3 (t) \int_{(X(t) + 2 \hat Q)\times \R^3} v^2 f(t,z)\, dz\\
&\leq & C r^{-1} a^3 (t) \int_S v^2 f(t,z)\, dz \\
&\leq & C r^{-1}.
\eeas
Combining the estimates for $I_1$, $I_2$, and $I_3$ we obtain
\beas
&&\n{a(t) V(t) - a(t - \Delta) V(t - \Delta)}\\
&&\qquad \qquad
\leq C \left( 1 + a(t) p^{4/3} + a^2(t) r \ln \frac{2 P(t)}{p} +
r^{-1} \Delta^{-1} \right) \Delta
\eeas
for
\[
\Delta = \Delta (t) = \min \left \{t, \frac{1}{2 \dot a(0)}, \frac{1}{4 P(t)},
\frac{p}{5 C_2 P(t)^{4/3}} \right\}
\]
Now choose $r= a^{-1}(t) P(t)^{1/2} \ln^{-1/2} (2 P(t)/p) $ and
$p=P(t)^{1/3}$. Then for $t$ close enough to $T$, say $t\geq t^\ast$, we have
$\Delta (t) = \frac{1}{5 C_2P(t)}$ and
\[
\n{a(t) V(t) - a(t - \Delta) V(t - \Delta)} \leq C a(t) P(t)^{1/2 + \epsilon}
\Delta (t)
\]
for any $\epsilon >0$. Defining $t_0 := t$ and $t_{i+1} := t_i - \Delta (t_i)$
as long as $t_i > t^\ast$ and observing that $\Delta (t_i) \geq \Delta (t_0)$
we see that there exists $k \in \N$ such that
$t_{k + 1} < t^\ast \leq t_k$. Thus
\beas
\n{a(t) V(t)} &\leq& a(t^\ast) \n{V(t^\ast)} + C a(t) P(t)^{1/2 + \epsilon}
\sum_{i=0}^k \Delta (t_i) \\
&\leq& C a(t) (1+t) P(t)^{1/2 + \epsilon},\ t\in [t^\ast, T[.
\eeas
We conclude that
\[
P(t) \leq C (1+t) P(t)^{1/2 + \epsilon}, \ t \in [0,T[,
\]
which proves the asymptotic estimate for $P(t)$ and by Thm.\ \ref{locex}
shows that $T=\infty$.

Let us now consider Case B: Assume that $T < T_a$. Then
$c_1 \leq a(t) \leq c_2$ and $\n{\dot a (t)} \leq c_3$ for all
$t \in [0,T[$ with positive constants $c_1, c_2, c_3$.
Now using exactly the same estimates as above we see that
\[
\n{a(t) V(t) - a(t - \Delta) V(t - \Delta)} \leq C P(t)^{1/2 + \epsilon}
\Delta (t)
\]
for $t$ close enough to $T$ and $\Delta (t):= \frac{1}{C P(t)}$.
This yields the same estimate for $P(t)$ on the interval $[0,T[$
as in Case A, in contradiction to Thm.\ \ref{locex}.
Thus, $T=T_a$ also in Case B, and the proof is complete.
\prfe

\end{document}